\documentclass[10pt,conference,english]{IEEEtran}
\usepackage[utf8]{inputenc}
\usepackage[english]{babel}
\usepackage[T1]{fontenc}

\usepackage{amsfonts}
\usepackage{amsmath}
\usepackage{amssymb}
\usepackage{amsthm}
\usepackage{bm}

\usepackage{tikz}
\usepackage{graphicx}
\usepackage{color}
\usepackage{pgfpages}

\newtheorem{proposition}{Proposition}
\newtheorem{lemma}{Lemma}
\newtheorem{definition}{Definition}

\newcommand{\graph}{\mathcal{G}}
\newcommand{\vertexv}{v}

\newcommand{\vertexe}{{\bf e}}

\newcommand{\vertices}{\mathcal{V}}
\newcommand{\edges}{\mathcal{E}}
\newcommand{\blackhole}{\omega}
\newcommand{\signal}{{\bf x}}

\title{Neighborhood-Preserving Translations on Graphs}

\author{Nicolas Grelier, Bastien Pasdeloup, Jean-Charles Vialatte and Vincent Gripon\\Télécom Bretagne, France\\name.surname@telecom-bretagne.eu}

\begin{document}

\maketitle

\begin{abstract}
  In many domains (e.g. Internet of Things, neuroimaging) signals are naturally supported on graphs.
  These graphs usually convey information on similarity between the values taken by the signal at the corresponding vertices.
  An interest of using graphs is that it allows to define ad hoc operators to perform signal processing.
  Among them, ones of paramount importance in many tasks are translations.
  In this paper we are interested in defining translations on graphs using a few simple properties.
  Namely we propose to define translations as functions from vertices to adjacent ones, that preserve neighborhood properties of the graph.
  We show that our definitions, contrary to other works on the subject, match usual translations on grid graphs.
\end{abstract}

\section{Introduction}

Graph signal processing proposes frameworks to define harmonic operators on domains characterized by a graph.
Using analogies with discrete Fourier calculus, it is thus possible to define ad hoc operators on the frequency domains including convolutions and wavelets.
Applications are numerous, ranging from compression to learning and may be applied in domains where each data point (typically a real valued vector) can be seen as scalars distributed over a network (typically characterized by the adjacency matrix of a graph).

Of particular interest are translation operators, for they underlie many others (e.g. convolutions).
A motivating example is extending convolutional neural networks to graph signals, thus making it possible to identify a same object at different locations in a graph~\cite{bruna2013spectral,vialatte2016generalizing}.
Another one is the ability to identify moving patterns in brain imaging to obtain better models for causal connectivity~\cite{friston2003dynamic}.

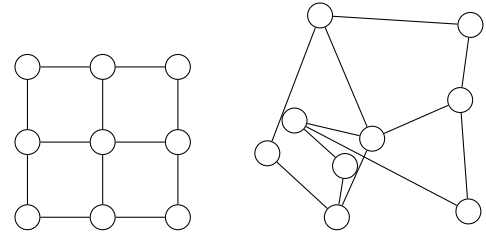
\begin{figure}[!h]
  \begin{center}
    \begin{tabular}{ccc}    
        \begin{tikzpicture}
          \foreach \x in {1,...,3}{
            \foreach \y in {1,...,3}{
              \node[draw, circle](\x\y) at (\x,\y) {};
            }
          }
          \foreach \x in {1,...,3}{
            \path[]
            (\x 1) edge (\x 2)
            (\x 2) edge (\x 3)
            (1\x) edge (2\x)
            (2\x) edge (3\x)
            ;
          }
          
        \end{tikzpicture}
      & &
        \begin{tikzpicture}
          \pgfmathsetseed{23}
          \foreach \x in {1,...,3}{
            \foreach \y in {1,...,3}{
              \node[draw, circle](\x\y) at (0.17*\x+1.5*rand,1.5*rand) {};
            }
          }
          \foreach \x in {1,...,3}{
            \path[]
            (\x 1) edge (\x 2)
            (\x 2) edge (\x 3)
            (1\x) edge (2\x)
            (2\x) edge (3\x)
            ;
          }

        \end{tikzpicture}
    \end{tabular}
  \end{center}
  \caption{Example of the same graph depicted twice with different vertices locations, emphasizing why it is not trivial to define translations on graphs that match usual ones on regular domains.}
  \label{fig:identiques}
\end{figure}

Several definitions of translations relying on spectral transforms have been proposed~\cite{shuman2013emerging, girault2015stationary}.
However, the obtained operators do not match usual translations on regular domains.
This finding is not surprising considering that in a graph vertices are not localized and only interconnections are considered.
This is depicted in Figure~\ref{fig:identiques} where the same graph has been represented twice but with different vertices locations.

Fundamentally, we are interested in the following question: is it possible to propose novel generic definitions of translations on graphs that ensure obtained operators match usual ones on regular domains?
We propose simple such definitions and prove they match usual translations on some regular domains.
Interestingly, these definitions mainly require preservation of neighborhoods.

The outline of the paper is as follows.
In Section 2 we introduce related work and definitions.
Section 3 contains the main results and proofs.
Section 4 is a conclusion. 

\section{Definitions and Related Work}
Throughout this paper we consider a \emph{graph} $\graph = \langle \vertices, \edges \rangle$ where $\vertices$ is the set of vertices and $\edges \subset \vertices \times \vertices$ is the set of edges.
We only consider graphs that are simple ($\forall \vertexv \in \vertices, (\vertexv,\vertexv) \not \in \edges$) and symmetric ($\forall \vertexv,\vertexv' \in \vertices, (\vertexv,\vertexv')\in \edges \Rightarrow (\vertexv',\vertexv)\in \edges$).
By abusing notations we make no distinction between edges $(\vertexv,\vertexv')$ and $(\vertexv',\vertexv)$.
We denote the set of integers between $a$ and $b$ (both included) $[\![a,b]\!]$.

\subsection{Grid Graphs}
We are particularly interested in grid graphs. Specifically, we consider cyclic grid graphs and noncyclic ones.
To define them, let us consider an integer $d \in \mathbb{N}^*$, that is the dimension of the grid graph.
We denote by $\left(\ell_i\right)_{1\leq i \leq d}$ the length of each dimension and consider that $\forall i \in [\![1,d]\!], \ell_i \geq 5$.

\begin{definition}[\emph{Cyclic grid graphs}] \normalfont
To define cyclic grid graphs, we introduce the Cartesian product group $\langle \vertices \triangleq \left(\mathbb{Z}/\ell_1\mathbb{Z}\right) \times \left(\mathbb{Z}/\ell_2\mathbb{Z}\right) \times \dots \times \left(\mathbb{Z}/\ell_d\mathbb{Z}\right), + \rangle$.
Elements in $\vertices$ are $d$-dimensional vectors.
We denote the canonical basis vectors $\left(\vertexe_i\right)_{1\leq i \leq d}$ such that $\vertexe_i[j] = 1$ if $j = i$ and $0$ otherwise.
We consider that two elements ${\bf v}_1, {\bf v}_2$ of $\vertices$ are \emph{neighbors} if ${\bf v}_1 - {\bf v}_2 = \pm \vertexe_i$.
Then the cyclic grid graph with parameters $d$ and $\left(\ell_i\right)_{1\leq i \leq d}$ is the graph $\graph = \langle \vertices, \edges \rangle$ such that $({\bf v}_1,{\bf v}_2) \in \edges$ if and only if ${\bf v}_1$ and ${\bf v}_2$ are neighbors.
\end{definition}

\begin{definition}[\emph{Noncyclic grid graphs}] \normalfont
The (noncyclic) grid graph with parameters $d$ and $\left(\ell_i\right)_{1\leq i \leq d}$ is the graph $\graph = \langle \vertices, \edges \rangle$ such that $\vertices = [\![0,\ell_1-1]\!] \times [\![0,\ell_2-1]\!] \times \dots \times [\![0,\ell_d-1]\!] $ and $({\bf v}_1,{\bf v}_2) \in \edges \Leftrightarrow d_{\text{taxi}}({\bf v}_1, {\bf v}_2)=1$ where $d_{\text{taxi}}$ is the taxicab distance.
\end{definition}

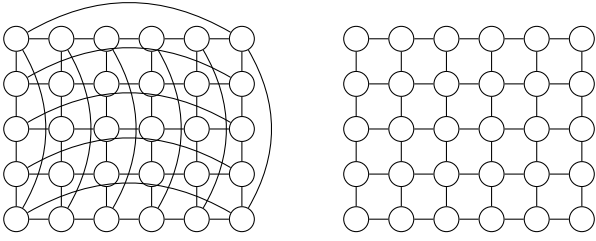
\begin{figure}
  \begin{center}
    
    \begin{tabular}{ccc}
        \begin{tikzpicture}[scale=0.6]
          \tikzstyle{every node} = [draw, circle];
          \foreach \i in {0,1,2,3,4,5}{
            \foreach \j in {0,1,2,3,4}{
              \node(\i\j) at (\i,\j) {};
            }
          }
          \path[]
          \foreach \i in {0,1,2,3,4,5}{
            \foreach \j/\jj in {0/1,1/2,2/3,3/4}{
              (\i\j) edge (\i\jj)
            }
          }
          \foreach \i/\ii in {0/1,1/2,2/3,3/4,4/5}{
            \foreach \j in {0,1,2,3,4}{
              (\i\j) edge (\ii\j)
            }
          }
          \foreach \i in {0,1,2,3,4,5}{
            (\i4) edge[bend left] (\i0)
          }
          \foreach \j in {0,1,2,3,4}{
            (0\j) edge[bend left] (5\j)
          }
          ;
        \end{tikzpicture}
      & &
              \begin{tikzpicture}[scale=0.6]
          \tikzstyle{every node} = [draw, circle];
          \foreach \i in {0,1,2,3,4,5}{
            \foreach \j in {0,1,2,3,4}{
              \node(\i\j) at (\i,\j) {};
            }
          }
          \path[]
          \foreach \i in {0,1,2,3,4,5}{
            \foreach \j/\jj in {0/1,1/2,2/3,3/4}{
              (\i\j) edge (\i\jj)
            }
          }
          \foreach \i/\ii in {0/1,1/2,2/3,3/4,4/5}{
            \foreach \j in {0,1,2,3,4}{
              (\i\j) edge (\ii\j)
            }
          }
          ;
        \end{tikzpicture}
    \end{tabular}
    
  \end{center}
  \caption{Example of a cyclic grid graph (left) and of a noncyclic grid graph (right), both of them with parameters $d = 2$ and $(\ell_1, \ell_2) = (6,5)$.}
  \label{fig:gridexample}
\end{figure}

Figure~\ref{fig:gridexample} depicts an example of a cyclic grid graph and a noncyclic grid graph, both with parameters $d = 2$ and $(\ell_1, \ell_2) = (6,5)$.

\subsection{Signals on Graphs}
Considering vertices to be indexed from 1 to $|\vertices|$, where $|\cdot|$ denotes the cardinality operator, a graph is characterized by a $|\vertices|$-dimensional binary square matrix ${\bf W}$ such that ${\bf W}[i,j] = {\bf W}[j,i] = 1$ if and only if the $i$-th vertex is connected through an edge to the $j$-th vertex (${\bf W}$ is an adjacency matrix associated with $\mathcal{G}$).

A signal $\signal$ on $\graph$ is a vector in $\mathbb{R}^{|\vertices|}$.
It can be seen as a collection of scalars associated with each vertex in the graph.
We are interested in defining a translation of $\signal$ on $\graph$.

\subsection{Related Work}

In~\cite{shuman2013emerging}, the authors propose to define a convolution operator first.
To this end, they use the Laplacian matrix of the graph, defined as ${\bf L} = {\bf D} - {\bf W}$, where ${\bf D}$ is the diagonal matrix such that ${\bf D}[i,i]$ is the number of vertices the $i$-th vertex is connected to.
Since ${\bf W}$ is symmetric, ${\bf L}$ also is, and it can be decomposed as ${\bf L} = {\bf U} {\bf \Lambda} {\bf U}^\top$, where ${\bf U}$ is an orthonormal matrix, ${\bf \Lambda}$ is a diagonal matrix and $\cdot ^\top$ denotes the transpose operator.
Then they define $\hat{\signal} \triangleq {\bf U}^\top \signal$, called the graph Fourier transform of $\signal$, and $\signal \triangleq {\bf U} \hat{\signal}$, termed the inverse graph Fourier transform of $\hat{\signal}$.
The authors introduce a convolution operator for two signals $\signal_1$ and $\signal_2$ as ${\bf U} ({\bf U}^\top \signal_1 \odot {\bf U}^\top \signal_2)$, where $\odot$ denotes the elementwise product of vectors.
They particularize the translation by convolving $\signal$ with a signal $\vertexe_\vertexv$ in the canonical basis.
Thus, a translation is not defined by a ``shift'' but by a destination vertex $\vertexv$.

In~\cite{girault2015stationary}, the authors focus on defining an isometric operator with respect to the $\ell_2$-norm.
The translation of a signal they propose consists in multiplying the signal by some matrix which is an exponential of an imaginary diagonal matrix.

Finally, some definitions of translation (or ``shift'') exist for powers of a ring graph~\cite{Ekambaram} for which it is quite straightforward.

As we already mentioned in the introduction, the obtained operators are not consistent with usual translations when applied on regular graphs such as grid graphs, which motivates the definitions introduced in the following subsection.

\subsection{Proposed Definitions}

\begin{definition}[\emph{Perfect graphical translation}] \normalfont
We say that an application $f$ from $\vertices$ to $\vertices$ is a perfect graphical translation of $\graph$ if:
\begin{enumerate}
	\item $f$ is bijective;
	\item for all vertex $\vertexv$ in $\vertices$, $f(\vertexv)$ is a neighbor of $\vertexv$;
	\item for all couple $(\vertexv_1,\vertexv_2) \in \vertices^2$,  $(\vertexv_1,\vertexv_2) \in \edges \Leftrightarrow (f(\vertexv_1),f(\vertexv_2)) \in \edges$.
\end{enumerate}
\end{definition}

Perfect graphical translations are hard to obtain in practice (examples of graphs that admit one are few).
In order to be able to define translations on any graph, we also propose a generalized definition of graphical translations.
To do so, we first define a \emph{black hole} $\blackhole\not\in \vertices$ onto which we can map some vertices.
Let $f$ be an application from $\vertices$ to $\vertices \cup \{ \blackhole\}$.
We denote $\vertices_{0,f} = f^{-1}\left(\vertices\right)$.

\begin{definition}[\emph{Candidate graphical translation}] \normalfont
We say that an application $f$ from $\vertices$ to $\vertices \cup \{ \blackhole\}$ is a candidate graphical translation if:
\begin{enumerate}
	\item $f_{|\vertices_{0,f}}$ is injective;
	\item for all vertex $\vertexv$ in $\vertices_{0,f}$, $f(\vertexv)$ is a neighbor of $\vertexv$;
	\item for all couple $(\vertexv_1,\vertexv_2) \in \vertices_{0,f}$,  $(\vertexv_1,\vertexv_2) \in \edges \Leftrightarrow (f(\vertexv_1),f(\vertexv_2)) \in \edges$.
\end{enumerate}
\end{definition}

\begin{definition}[\emph{Generalized graphical translation}] \normalfont
We say that an application $f$ from $\vertices$ to $\vertices \cup \{ \blackhole\}$ is a (generalized) graphical translation if:
\begin{enumerate}
	\item $f$ is a candidate graphical translation;
	\item $\forall \vertexv \in \vertices_{0,f}$, for every $g$ candidate graphical translation such that $g(\vertexv)=f(\vertexv)$, $|\vertices_{0,g}| \leq |\vertices_{0,f}|$.
\end{enumerate}
\end{definition}

By construction of grid graphs, we have in mind what translations should be, as pointed out in the following definition. 

\begin{definition}[\emph{Geometrical translation}] \normalfont
We say that an application $f$ from $\vertices$ to $\vertices \cup \{ \blackhole\}$ is a geometrical translation on a grid graph $\graph = \langle \vertices, \edges \rangle$ if there is ${\bm \delta} = \pm \vertexe_i$ such that \[\forall {\bf v} \in \vertices, f({\bf v}) = \left\{
\begin{array}{ll}
  {\bf v} + {\bm \delta} & \text{if } {\bf v} + {\bm \delta} \in \vertices \\
  \blackhole & \text{otherwise}
\end{array}
\right. .\]
\end{definition}

Note that our definition of geometrical translations only considers elementary ones, where the shift length is minimum.
More generic translations can be obtained by composing elementary ones.

We are interested in showing that:
\begin{itemize}
\item On cyclic grid graphs, graphical translations are geometrical translations, and they are perfect;
\item On noncyclic grid graphs, graphical translations are geometrical translations.
\end{itemize}

\section{Main Results}

\subsection{Cyclic Grid Graphs}

Let us begin with cyclic grid graphs.
First we point out that if a cyclic grid graph admits perfect graphical translations then they are the only possible graphical translations.
We then proceed in two steps: first we show that perfect graphical translations are geometrical translations, then the converse.

\begin{lemma} [\emph{Contamination lemma}] \normalfont
Let $f$ be a perfect graphical translation on a cyclic grid graph $\graph = \langle \vertices, \edges \rangle$ with parameters $d$ and $\left(\ell_i\right)_{1 \leq i \leq d}$.
Let ${\bf v}$ be in $\vertices$ and consider ${\bm \delta} = f({\bf v}) - {\bf v}$, then $\forall {\bf w} \in \vertices$ neighbor of ${\bf v}$, $f({\bf w}) = {\bf w} + {\bm \delta}$.
\label{lemma:conta}
\end{lemma}

\begin{IEEEproof}
  Let us consider ${\bf w}$ neighbor of ${\bf v}$.

  Let us show that $f({\bf v} - {\bm \delta}) = {\bf v}$.
  As a matter of fact, Property 2) of perfect graphical translations gives us that $f({\bf v} - {\bm \delta})= ({\bf v} - {\bm \delta}) + {\bm \epsilon}$, with ${\bm \epsilon} = \pm \vertexe_i$.
  Property 3) forces that ${\bm \epsilon} = \pm {\bm \delta}$.
  Moreover we have $\forall i \in [\![1,d]\!], \ell_i \geq 5$, thus $f({\bf v} - {\bm \delta})={\bf v}$.

  Let ${\bf w}$ be in $\vertices - \{{\bf v} + {\bm \delta}, {\bf v} - {\bm \delta}\}$, such that ${\bf v}$ and ${\bf w}$ are neighbors.
  Property 3) gives us that $(f({\bf w}),f({\bf v})) \in \edges$.
  Moreover, Property 2) gives us that $({\bf w},f({\bf w})) \in \edges$, so $f({\bf w}) = {\bf v}$ or $f({\bf w}) = {\bf w} + {\bm \delta}$.
  Furthermore $f({\bf w}) \neq {\bf v}$ because $f$ is a bijection and $f({\bf v} - {\bm \delta}) = {\bf v}$.
  We conclude: $f({\bf w}) = {\bf w} + {\bm \delta}$.

Let ${\bf w} = {\bf v} +{\bm \delta}$, then we have $f({\bf w}) = {\bf w} \pm {\bm \delta}$ because of the precedent result applied to ${\bf w}$.
Furthermore, as proven above, we have $f({\bf v} -{\bm \delta})= {\bf v}$. Thus $f({\bf w}) = {\bf w} +{\bm \delta}$ because of the bijectivity of $f$. 
\end{IEEEproof}

\begin{proposition} \normalfont
  Let $f$ be a perfect graphical translation on a cyclic grid graph, then $f$ is a graphical translation.
\end{proposition}

\begin{IEEEproof}
  Let us denote ${\bm \delta} = f({\bf 0})-{\bf 0}= f({\bf 0})$ and let ${\bf v} \in \vertices$.
  As a cyclic grid graph is connected, there is a path from ${\bf 0}$ to ${\bf v}$.
  By propagating the contamination lemma along the path, neighbor after neighbor, we obtain that $f({\bf v}) = {\bf v} + {\bm \delta}$.
\end{IEEEproof}

For the converse result, the proof is straightforward and therefore omitted.

\begin{proposition} \normalfont
Let $f$ be a geometrical translation on a cyclic grid graph $\graph$, then $f$ is a perfect graphical translation on $\graph$.
\end{proposition}

Note that a direct consequence of these proofs is that there are as many perfect graphical translations as there are neighbors for a given vertex in a cyclic grid graph.

\subsection{Noncyclic Grid Graphs}

We then proceed with noncyclic grid graphs.
We restrict our proofs to subfamilies such graphs.
Namely, we only consider graphs with parameters $d$ and $\left(\ell_i\right)_{1 \leq i \leq d}$ such that:
\[
\forall i \in [\![1,d-1]\!],\ell_i \geq \left(2 \prod_{k=i+1}^{d} \ell_k\right)+2.
\]
Moreover, we force $\ell_d \geq 3$.

Let us first point out that the function $f_\blackhole$:
\[
f_\blackhole: \left\{
\begin{array}{rcl}
  \vertices &\to& \vertices \cup \{\blackhole\}\\
  {\bf v} &\mapsto & \blackhole
\end{array}
\right.
\]
is a graphical translation.

More generally, a graph may admit several graphical translations $f$ with various sizes of $\vertices_{0,f}$. When applicable, we refer to them as $c$-graphical translations where $c = |\vertices_{0,f}|$.




\begin{lemma} \normalfont
  Let $f$ be a candidate graphical translation of any graph $\graph = \langle \vertices, \edges \rangle$.
  For any vertex ${\bf v} \in \vertices$, the sequence $\left(f^n({\bf v})\right)_n$ is either periodic or finite (in which case the last element is $\blackhole$).
  \label{lemma:blackhole}
\end{lemma}

\begin{IEEEproof}
  Let us discuss whether there exists some $n$ such that $f^n({\bf v}) = \blackhole$ or not.
  If it exists then we are done.
  Otherwise, as $\vertices$ is finite, there exists $p,q$, with $p < q$, such that $f^p({\bf v})=f^q({\bf v})$.
  Moreover, $f$ being injective, we conclude that $f^{q-p}({\bf v}) = {\bf v}$.
\end{IEEEproof}

\emph{Remark:} $|\vertices|$-graphical translations -- which are also perfect graphical translations -- split the graph into cycles.

In order to ease reading of the following results, we introduce the definition of a slice in a grid graph.
\begin{definition}[\emph{Grid graph slice}] \normalfont
We call grid graph slice $(a,i)^\bot$ a set of vertices that share one coordinate of value $a$ at dimension $i$.
\end{definition}

\emph{Remark:} the cardinal of a a slice $(a,i)^\bot$ is $\prod_{k=1, k\not=i}^{d} \ell_k$.

\begin{lemma} \normalfont
  Let us consider a noncyclic grid graph $\graph = \langle \vertices, \edges \rangle$ with parameters $d$ and $(\ell_i)_{1 \leq i \leq d}$, and $f$ a $c$-graphical translation with the largest $c$.
  Then we have $|\vertices_{0,f}| \geq (\ell_1-1) \prod_{k=2}^{d} \ell_k$.
  \label{lemma:bound}
\end{lemma}
\begin{IEEEproof}
Let us consider $g$ to be the geometrical translation by $\vertexe_1$.
Then $\forall {\bf v} \in \vertices, g({\bf v})= \blackhole \Leftrightarrow {\bf v} \in (\ell_1,1)^\bot$.
Since we have $|\vertices_{0,g}| \leq |\vertices_{0,f}|$, we conclude $|\vertices_{0,f}| \geq (\ell_1-1) \prod_{k=2}^{d} \ell_k$.
\end{IEEEproof}

\begin{lemma} \normalfont
  Consider a union of two adjacent slices $\mathcal{S}$ of a noncyclic grid graph $\graph$.
  Consider $f$ to be a $c$-graphical translation with the largest $c$.
  If there exists some vertex ${\bf v} \in \vertices$ such that $\left(f^n({\bf v})\right)_n$ is periodic, then $\mathcal{S} \not\subset \vertices_{0,f}$. 
  \label{lemma:cycle}
\end{lemma}

\begin{IEEEproof}
  Let us consider a vertex ${\bf v}$ such that the corresponding sequence $\left(f^n({\bf v})\right)_n$ is periodic.
  We then necessarily have $n$ such that $f^{n+2}({\bf v}) - f^{n+1}({\bf v}) \neq f^{n+1}({\bf v}) - f^{n}({\bf v})$, since by contradiction we would obtain a nonperiodic sequence.
  Note that the period of the sequence cannot be $2$ since it would lead to at least $2 \prod_{k=2}^{d} \ell_k$ elements in $\vertices - \vertices_{0,f}$ using a contamination principle as in Lemma~\ref{lemma:conta}, thus contradicting Lemma~\ref{lemma:bound}.

  We thus obtain a ``turn'' in the sequence of states.
  A turn necessarily leads to a periodic sequence of $4$ vertices.
  As a matter of fact, let us denote by ${\bf w} = f^n({\bf v}) + f^{n+2}({\bf v}) - f^{n+1}({\bf v})$, since ${\bf w}$ and $f^n({\bf v})$ are neighbors, their images must also be, forcing $f({\bf w})=f^n({\bf v})$.
  Then $f^{n+3}({\bf v}) = {\bf w}$.
  We obtain $\forall n, f^{n+4}({\bf v}) = f^{n}({\bf v})$.
  
  We conclude that $\forall n \in [\![ 0, 3]\!], 2f^{n}({\bf v}) -f^{n+1}({\bf v}) \not\in \vertices_{0,f}$.
  At least one of these vertices is in $\mathcal{S}$ since $\forall i,\ell_i\geq 3$.
\end{IEEEproof}

\begin{lemma} \normalfont
  Let us consider $f$ to be a $c$-graphical translation with the largest $c$.
  Then there exists $m$ such that $(m,1)^\bot \subset \vertices_{0,f}$ and $(m+1,1)^\bot \subset \vertices_{0,f}$.
  Moreover, $f\left((m,1)^\bot \cup (m+1,1)^\bot\right) \not \subset (m,1)^\bot \cup (m+1,1)^\bot$.
  \label{lemma:adjacent}
\end{lemma}

\begin{IEEEproof}
  Lemma~\ref{lemma:bound} shows us that $|\vertices -\vertices_{0,f}| \leq \prod_{k=2}^{d} \ell_k$.
  Also we have $\ell_1 \geq (2 \prod_{k=2}^{d} \ell_k)+2$.
  Thus because of the pigeonhole principle, there exists ${\bf v}$ such that $(m,1)^\bot \subset \vertices_{0,f}$ and $(m+1,1)^\bot \subset \vertices_{0,f}$.

  Let us discuss two cases depending on the image of some vertex ${\bf v}$ in $(m,1)^\bot$, which is not empty. 
  
  1) $\exists n \in \mathbb{N}$, such that $f^n({\bf v}) \not\in (m,1)^\bot \cup (m+1,1)^\bot$.
  
  2) We iterate $f$ to obtain the sequence $\left(f^n({\bf v})\right)_n$.
  By Lemma~\ref{lemma:blackhole}, we know that either this sequence ends with $\blackhole$ in which case we are done, or it is periodic and all its elements are in $(m,1)^\bot \cup (m+1,1)^\bot$.
  In the latter case, we conclude with Lemma~\ref{lemma:cycle}.
\end{IEEEproof}

\begin{proposition} \normalfont
  Let us consider $f$ to be a $c$-graphical translation with the largest $c$.
  Then $f$ is the geometrical translation by $\vertexe_1$ or $-\vertexe_1$.
  \label{prop:majoraxe}
\end{proposition}

\begin{IEEEproof}
  Let us apply Lemma~\ref{lemma:adjacent} and suppose without loss of generality that $\exists m, j_2,\dots,j_d$ such that $f((m,j_2,\dots,j_d))=(m-1,j_2,\dots,j_d)$.
  Using the same principle as in Lemma~\ref{lemma:conta}, we obtain $\forall {\bf v} \in (m,1)^\bot\cup (m+1,1)^\bot, f({\bf v})={\bf v}-\vertexe_1$.
  Considering Lemma~\ref{lemma:blackhole} and starting at any vertex ${\bf v}$ in $(m+1,1)^\bot$, we obtain that $\left(f^n({\bf v})\right)_n$ is either periodic or contains $\blackhole$.
  We then observe that it cannot be periodic since all elements in $(m,1)^\bot$ are already images of elements in $(m+1,1)^\bot$.
  We conclude that the subset of vertices in the slices $(0,1)^\bot \cup \dots \cup (m-1,1)^\bot$ contains at least $\prod_{k=2}^{d} \ell_k$ elements in $\vertices - \vertices_{0,f}$.

  Since we supposed $f$ to be a $c$-graphical translation with the largest $c$, and using Lemma~\ref{lemma:bound}, we conclude that all vertices in $(m,1)^\bot \cup \dots \cup (\ell_1-1,1)^\bot$ are in $\vertices_{0,f}$.
  Using the same principle as in Lemma \ref{lemma:conta}, we conclude that $\forall {\bf v} \in (m,1)^\bot\cup\dots \cup (\ell_1-1,1)^\bot, f({\bf v}) = {\bf v} - \vertexe_1$.

  Then let us suppose by contradiction that $f$ is not a geometrical translation by $-\vertexe_1$.
  And let us look at the largest $k < m$ (we thus have $k > 0$) such that $(k,1)^\bot$ contains an element in $\vertices-\vertices_{0,f}$.
  We denote it by ${\bf v}$.
  Because of Properties 2) and 3), we obtain that ${\bf v} - \vertexe_1$ is not the image of any other vertex.
  We conclude that we have at least $1 + \prod_{k=2}^{d} \ell_k$ elements in $\vertices-\vertices_{0,f}$.
  Lemma~\ref{lemma:bound} concludes.
\end{IEEEproof}

\begin{proposition} \normalfont
  Graphical translations on cyclic grid graphs are geometrical translations and $f_\blackhole$.
\end{proposition}

\begin{IEEEproof}
  Let us consider a cyclic grid graph with parameters $d$ and $\left(\ell_i\right)_i$.
  Proposition~\ref{prop:majoraxe} shows us that the $c$-graphical translations with the largest $c$ are geometrical translations by $\pm\vertexe_1$.
  As a consequence, the other graphical translations are such that $\forall {\bf v}, f({\bf v}) \neq {\bf v} \pm \vertexe_1$.
  Consider any slice $(m,1)^\bot$ and restrict our study to the induced subgraph, whose dimension is $d-1$.
   Considering one subgraph independently from the others, Proposition 3 gives us that the graphical translations on this subgraph are the geometrical translations by $\vertexe_2$ or $- \vertexe_2$. Those translations are $c'$-graphical translations with the largest $c'$, that is $c'=(\ell_2-1)\prod_{k=3}^{d}{\ell_k}$.
   Moreover, when considered jointly, the geometrical translations by $\vertexe_2$ or $- \vertexe_2$ do not add new vertices in $\vertices - \vertices_{0,f}$. We obtain a $(c' \ell_1)$-graphical translation on the initial graph.
   We repeat this process to obtain the result.
\end{IEEEproof}

\begin{figure}
  \begin{center}
    \begin{tikzpicture}[scale=0.86]
      \tikzstyle{every node}=[draw,circle];
      \node(00) at (0,0) {};
      \node(01) at (0,1) {};
      \node(02) at (0,2) {};
      \node(10) at (1,0) {};
      \node[fill=black](11) at (1,1) {};
      \node(12) at (1,2) {};
      \node(20) at (2,0) {};
      \node[fill=black](21) at (2,1) {};
      \node(22) at (2,2) {};
      \path[->]
      (00) edge (01)
      (01) edge (02)
      (02) edge (12)
      (12) edge (22)
      (22) edge (21)
      (20) edge (10)
      (10) edge (00);
      \draw[]
      (-0.5,-0.5) rectangle (2.5,2.5);
    \end{tikzpicture}
    ~
    \begin{tikzpicture}[scale=0.65]
      \tikzstyle{every node}=[draw,circle];
      \node[fill=black](00) at (0,0) {};
      \node(01) at (0,1) {};
      \node(02) at (0,2) {};
      \node(03) at (0,3) {};
      \node(10) at (1,0) {};
      \node(11) at (1,1) {};
      \node[fill=black](12) at (1,2) {};
      \node(13) at (1,3) {};
      \node(20) at (2,0) {};
      \node(21) at (2,1) {};
      \node[fill=black](22) at (2,2) {};
      \node(23) at (2,3) {};
      \node(30) at (3,0) {};
      \node(31) at (3,1) {};
      \node(32) at (3,2) {};
      \node(33) at (3,3) {};
      \path[->]
      (30) edge (20)
      (20) edge (10)
      (10) edge (00)
      (31) edge (21)
      (21) edge (11)
      (11) edge (01)
      (01) edge (02)
      (02) edge (03)
      (03) edge (13)
      (13) edge (23)
      (23) edge (33)
      (33) edge (32)
      (32) edge (31);
      \draw[]
      (-0.5,-0.5) rectangle (3.5,3.5);
    \end{tikzpicture}
  ~
    \begin{tikzpicture}[scale=0.52]
      \tikzstyle{every node}=[draw,circle];
      \foreach \x in {0,...,4}{
        \foreach \y in {0,...,4}{
          \node(\x\y) at (\x,\y) {};
        }
      }
      \node[fill=black] at (1,1) {};
      \node[fill=black] at (1,2) {};
      \node[fill=black] at (1,3) {};
      \node[fill=black] at (3,0) {};
      \node[fill=black] at (4,0) {};
      \path[->]
      (00) edge (01)
      (01) edge (02)
      (02) edge (03)
      (03) edge (04)
      (04) edge (14)
      (14) edge (24)
      (24) edge (23)
      (23) edge (22)
      (22) edge (21)
      (21) edge (20)
      (20) edge (10)
      (10) edge (00)
      (34) edge (33)
      (33) edge (32)
      (32) edge (31)
      (31) edge (30)
      (44) edge (43)
      (43) edge (42)
      (42) edge (41)
      (41) edge (40);
      \draw[]
      (-0.5,-0.5) rectangle (4.5,4.5);
    \end{tikzpicture}
  \end{center}
  \caption{Examples of 2-dimensional squared noncyclic grid graphs for which some graphical translations are not geometrical translations.
  Edges of the graph are not depicted.
  Graphical translations are depicted by arrows pointing vertices to their images through the considered graphical translation.
  Vertices pointing to $\blackhole$ are filled in black.}
  \label{fig:counterexamples}
\end{figure}
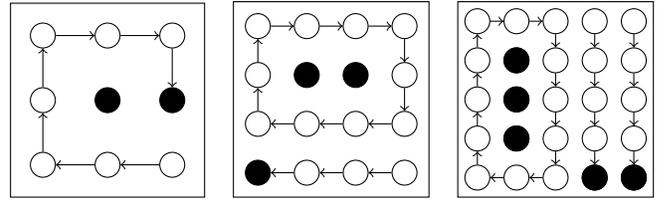

Generalizing these results to other noncyclic grid graphs does not seem trivial.
Figure~\ref{fig:counterexamples} depicts some examples of graphical translations obtained on 2-dimensional squared noncyclic grid graphs that are not geometrical translations.
Note that we were unable to find counterexamples for such graphs where the length of dimensions is at least 6.
Actually, we conjecture that any graphical translation on a noncyclic grid graph such that each dimension length is at least 6 is either a geometrical translation or $f_\blackhole$.

\begin{figure}
  \begin{center}
    \includegraphics[width=\linewidth]{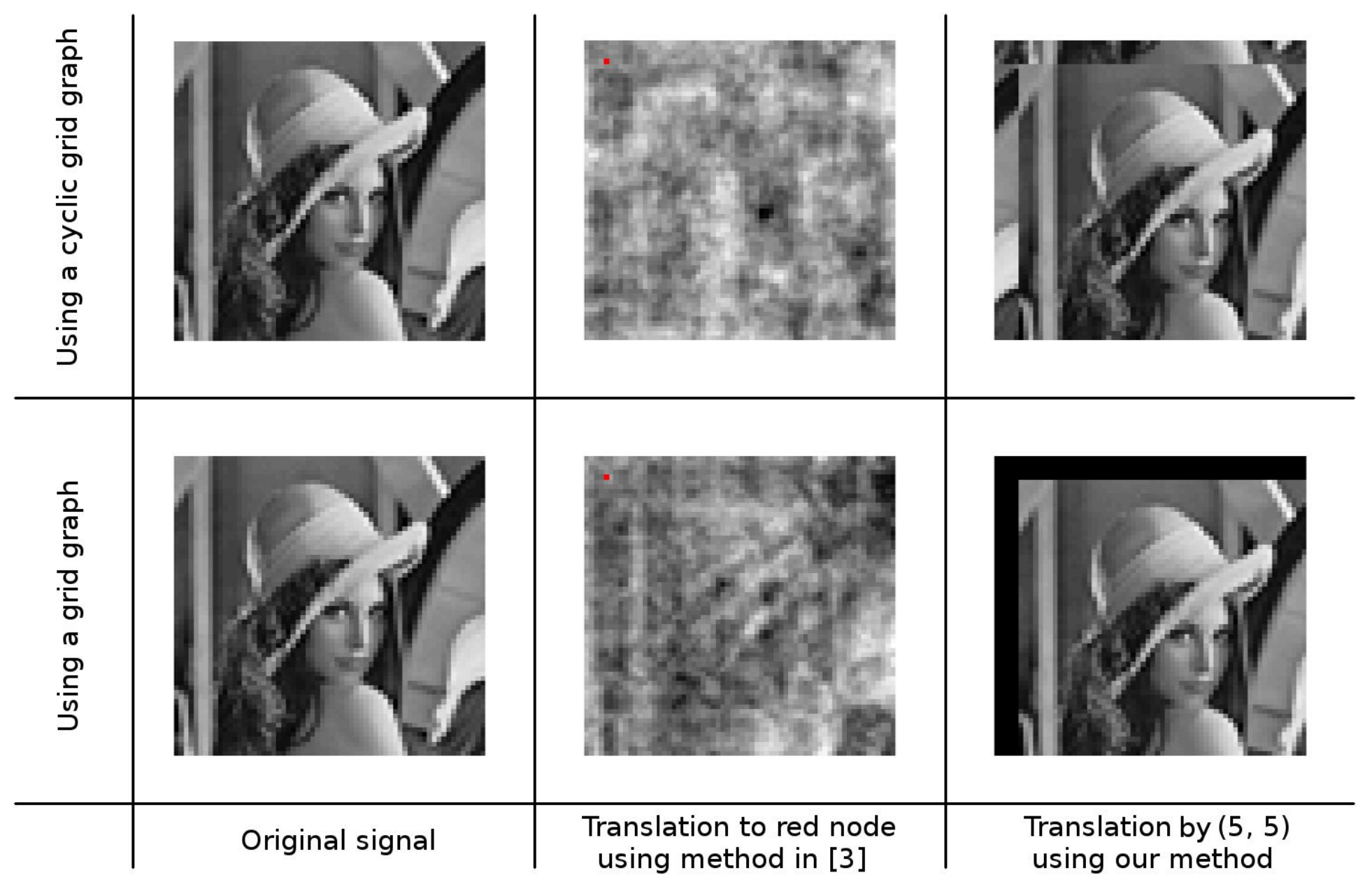}
  \end{center}
    \caption{Comparison of the effect of a translation as defined in~\cite{shuman2013emerging} with our neighborhood-preserving method.}
    \label{fig:lena}
\end{figure}

Finally, to illustrate our proposed translation on a graphical example, Figure~\ref{fig:lena} compares it to the translation of the classical $64 \times 64$ picture of Lena using the method introduced in~\cite{shuman2013emerging} as implemented in the GSP Toolbox~\cite{perraudin2014gspbox}.

\section{Conclusion}

We introduced a definition for translations on graphs that relies only on neighborhood preserving properties. When used on cyclic grid graphs, we proved that these definitions match usual translations. We also obtained results for some subfamilies of noncyclic grid graphs.
Interestingly, these definitions are simple and capture intuitions about translations. We believe that changing Property 2) should allow to extend to other types of isometries, including rotations.

We mainly focused our study on grid graphs, especially in order to obtain provable results. Finding translations on arbitrary graphs could prove challenging, considering the combinatorial explosion in the number of candidate graphical translations.
Future work include performing tests on distorted graphs, which will probably require modifying the definitions, as well as obtaining scalable algorithms to find graphical translations on large graphs.

\section*{Acknowledgements}
This work was funded in part by the European Research Council under the European Union's Seventh Framework Programme (FP7/2007 - 2013)/ERC grant agreement number 290901 and by the Labex CominLabs project ``Neural Communications''.

\bibliographystyle{IEEEtran}
\bibliography{bib}

\end{document}